\documentclass[aps,prl,twocolumn,superscriptaddress]{revtex4}
\usepackage{amssymb}
\usepackage{graphicx}
\usepackage{epsfig}
\usepackage{float}
\usepackage{amsmath}

\begin{document}

\title{The electronic structure of the Na$_x$CoO$_2$ surface}
\author{D. Pillay}
\affiliation{Code 6393, Naval Research Laboratory, Washington, D.C. 20375}
\author{M.D. Johannes}
\affiliation{Code 6393, Naval Research Laboratory, Washington, D.C. 20375}
\author{I.I. Mazin}
\affiliation{Code 6393, Naval Research Laboratory, Washington, D.C. 20375}

\begin{abstract} The idea that surface effects may play an important role in suppressing $e_g'$ Fermi surface 
pockets on Na$_x$CoO$_2$ $(0.333 \le x \le 0.75)$ has been frequently proposed to explain the discrepancy 
between LDA calculations (performed on the bulk compound) which find $e_g$' hole pockets present and ARPES 
experiments, which do not observe the hole pockets. Since ARPES is a surface sensitive technique it is 
important to investigate the effects that surface formation will have on the electronic structure of 
Na$_{1/3}$CoO$_2$ in order to more accurately compare theory and experiment. We have calculated the band 
structure and Fermi surface of cleaved Na$_{1/3}$CoO$_2$ and determined that the surface non-trivially affects 
the fermiology in comparison to the bulk. Additionally, we examine the likelihood of possible hydroxyl 
contamination and surface termination. Our results show that a combination of surface formation and 
contamination effects could resolve the ongoing controversy between ARPES experiments and theory. 
\end{abstract}

\maketitle One longstanding controversial issue surrounding Na$_x$CoO$_2$ is the discrepancy between between experimental 
measurements and theoretical predictions\cite{DJS00} concerning the small hole pockets along the $\Gamma-$K direction. Angle resolved 
photoemission spectra (ARPES) experiments have clearly shown the existence of a large a$_{1g}$ cylindrical hole on the Fermi 
surface which is consistent with LDA band structure calculations \cite{H-Y04,MZH+03,yang05,qian06}. However, the small hole pockets which are also present in 
band structure calculations due to the partially filled e$_g$' band are absent. Whether this should be interpreted as a failure 
of LDA to account for some crucial correlation effects or as a failure of ARPES to detect the bulk structure is a subject of 
much discussion.  

Dynamical mean field theory (DMFT) calculations that incorporate correlation effects on a local level and 
might be expected to bring theory and experiment into closer agreement, instead find that these only serve 
to {\it 
increase} the volume of the e$_g$' hole pockets \cite {ishida,georges,marianetti}. Nevertheless, since long range 
correlations such as ferromagnetic spin fluctuations are not treated by this method, one cannot rule out alterations to the Fermi surface 
caused by such correlations.

Surface effects, as a broadly defined category, have also been proposed as a possible reason for the absence of the pockets in 
ARPES measurements. By probing the Na 2$p$ core-level intensity as a function of photoelectron emission angle, it was clearly 
shown that strong differences between the surface and bulk exist and that photoemission experiments contain a large, if not 
dominant, amount of surface character \cite{arakane}. In particular, oxygen vacancies are common on oxide surfaces allowing hydroxyl 
terminations to occur through a combination of 
oxygen vacancies and a water background in UHV or simply 
hydrogen contaminants adsorbing on the surface \cite{Wendt:2006}. Indeed, many previous 
studies of oxide surfaces have found that it is nearly 
impossible to remove water and hydrogen contaminants in a UHV environment \cite{Brown:1999}.

It is therefore desirable to understand, from a theoretical point of 
view, the differences between the electronic structure of a cleaved surface layer and that of the bulk material and to 
investigate the role of common defects such as unintended adsorbates on the fermiology of NaCoO$_2$. 
In this Letter, we use density functional theory (DFT) to examine the band structure and Fermi surfaces of a cleaved 
Na$_{1/3}$CoO$_2$ surface.  We show that these differ non-trivially on a surface as compared to the bulk and are sensitive to 
many uncontrollable factors, such as Na ordering, surface contaminants, and defects.  Intriguingly, we find that the surface is 
extremely susceptible to electron donor adsorbates and that these strongly depress the e$_g$' hole pockets well beneath the 
Fermi energy, very similar to what has been observed in ARPES.

\textit{Calculational Methods} All calculations have been performed using a plane wave DFT code, the Vienna Ab-initio Simulation 
Package (VASP) \cite{vasp}. The projected augmented wave method (PAW)\cite{PAW1} and the generalized gradient approximation 
\cite{Perdew:1996} to the exchange correlation potential were used. The surface and bulk Na$_{x}$CoO$_{2}$ structures were modeled 
using a six stoichiometric layer slab (a 15{\AA } vacuum is added between successive slabs for surface calculations), so that both 
BZs are the same. For most calculations, we used a $\sqrt{3}\times\sqrt{3}$ hexagonal cell with three Co atoms per layer. The two 
top and bottom layers were relaxed while the middle two were held in their bulk relaxed positions. Ground state geometries were 
fully optimized using a Monkhorst Pack k-point mesh of 6$\times $4$\times $2. For total energy density of states (DOS) calculations, 
the Brillioun zone (BZ) integrations were performed on a 8$\times $8$\times $2 mesh.  For some calculations, an even larger 
$2\sqrt{3}x\sqrt{3}$ cell with six Co atoms per layer was required. Na can occupy either the sites atop Co (Na1) or at the center of 
an equilateral triangle formed by Co atoms at the vertices (Na2), or a combination of the two.  We have found that the energetics 
associated with Na2 ordering are most favorable in the $\sqrt{3}\times\sqrt{3}$ cell, so we present those results here.  This agrees 
well with experimental results from Roger {\it et. al} \cite{Roger:2007} who observe from neutron diffraction experiments that when 
x=1/3, Na prefers the Na2 position, whereas at high doping, multivacancy clusters form with partial occupation of Na1 sites.

In order for the stoichiometry of the bulk compound to be maintained and for the surface Co atoms to have the same valency as those 
in the bulk, the number of Na atoms on each of the two surfaces should be half the number between bulk layers.  For the $x$=1/3 
compound, this requires one Na for every six Co and cannot be achieved with only 3 Co atoms per layer.  We have therefore 
investigated an "over-stoichiometric" compound that contains one Na atom on each surface and an "under-stoichiometric" compound that 
has no Na on either surface.  The stoichiometric compound was also investigated by further lowering the symmetry (doubling the cell 
size).  Although we discuss the energetics of this calculation, the extensive backfolding of the BZ makes it impossible to visually 
interpret the Fermi surfaces.  We have investigated the stability of the under- or over-stoichiometric surfaces by calculating the 
phase separation energy difference $\frac {E_{1/3-}+E_{1/3+}}{2} \ge E_{1/3}$ where $E_{1/3-}$, $E_{1/3+}$ and $E_{1/3}$ represent 
the energies associated with the under-stoichiometric, overstoichiometric, and stoichiometric systems, respectively. We find this 
inequality to be true and therefore it is highly likely that the stoichiometric surface will not phase separate into areas that are 
either under- (Na depletion) or over- (Na excess) stoichiometric.

%\begin{equation}
%\frac {E_{1/3-}+E_{1/3+}}{2} \ge E_{1/3} \label{eqn1}\\
%\end{equation} 

We have adjusted the electron count in calculations 
of both under- and over-stoichiometric surfaces 
by shifting the Fermi energy appropriately to best mimic the most likely situation of a stoichiometric 
surface. As we discuss, this approximation is not perfectly representative because the Na atoms in some 
cases have a non-negligible effect on the surface Fermiology. It does, however, allow us to "bracket" expected
electronic structure modifications between over- and under- stoichiometric results. 

\begin{figure}[tbp]
\includegraphics{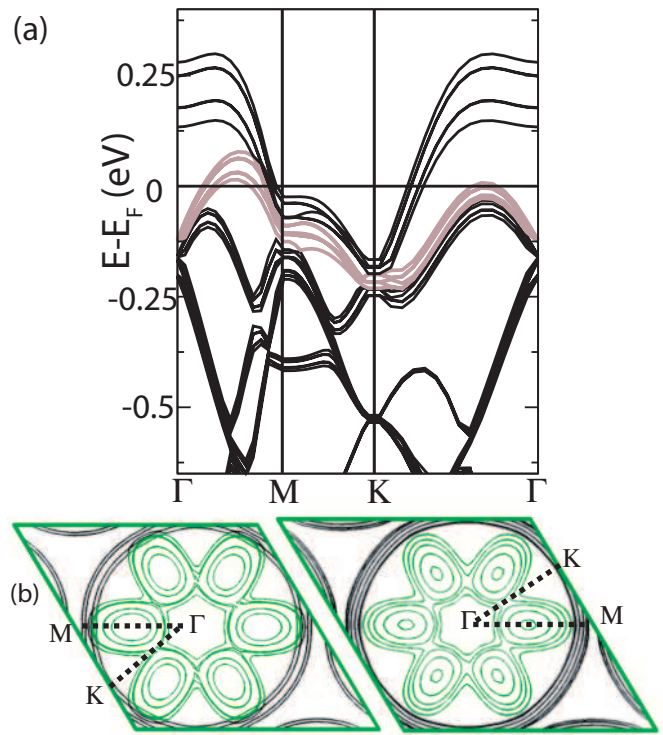}
\caption{(a) Band structure of bulk and (b)(left panel) Fermi surface of bulk 
and (right panel)overstoichiometric Na$_{1/3}$CoO$_2$. The light 
colored solid bands show e$_g$' character. The $a_{1g}$
bands create the large, dark (black) cylindrical pockets. The elliptical 
hole pockets are shown in light (green) and
arise from $e_g$' bands.}
\label{FS13}
\end{figure}

\textit{Surface Calculations} The bulk band structure and Fermi surfaces for Na$_{1/3}$CoO$_{2}$ in a 
six-layer $\sqrt{3}\times \sqrt{3}$ unit cell is shown in Figs. \ref{FS13}a and b.  There are six sets of 
$a_{1g}$ and $e_g$' Fermi surfaces (degeneracies prevent them from being individually 
distinguishable), corresponding to the six different layers. Because of BZ backfolding the two 
types of Fermi level crossings occur along the same high symmetry line, $\Gamma -$M, in Fig.\ref{FS13}a 
and the $e_g$'-derived small hole pockets are located inside the larger $a_{1g}$ hexagonal 
pockets in Fig.\ref{FS13}b. For the bulk compound, locating the Na atoms at the Na1 vs. the Na2 position 
has an imperceptible effect on the Fermi surface or band structure.  In Fig. \ref {FS13}b (right panel) 
we show the 
Fermi surface for the over-stoichiometric surface compound. We present the over-stoichiometric case 
because although the effect of Na is exaggerated it more closely mimics the stoichiometric case than neglecting it 
entirely. Moreover, the under-stoichiometric case also shows a noticeable 
decrease in the size of the $e_g$' hole pockets, though smaller than 
the over-stoichiometric case. The stoichiometric case, if it were possible to calculate it in this unit cell, would lie
somewhere between the two. In the over-stoichiometric case one of the $e_{g}$' pockets is 
clearly shrunken and another has been suppressed entirely. 
These can be clearly identified, using atom-based 
projection, as the surface-originated pockets. The corresponding surface a$_{1g}$ bands, identified in the 
same way, expand proportionately.  Thus, any technique that probes the surface electronic structure cannot
be expected to detect Fermi surfaces in agreement with bulk calculations and, additionally, surface 
sensitive results should not be extrapolated to represent the bulk electronic structure.  While the 
$e_{g}$' hole pockets are not fully eliminated by the existence of the surface, they are rather 
dramatically shrunken.  The origin of this effect is in the noticeably shortened Na-O distance (it decreases by 0.32 {\AA} at 
the surface). The Na$^+$ Coulomb field lowers the energy of the O
electron bands and, by increasing the energy separation between O and Co states, decreases Co-O bonding.  Although we do not show it
here, a projection of the O character of the Co-derived t$_{2g}$ bands shows that Co-O hybridization is stronger in the e$_g$'
symmetry parts of the band complex relative to the a$_{1g}$ parts and particularly strong at and near the part of the dispersion
that gives rise to the small e$_g$' fermi surfaces.  Thus, an energy lowering of the O states disproportionately affects these 
areas.

 In contrast to the bulk, the surface Fermi pockets (both 
small and large) are 
slightly sensitive to the Na position.  The Fermi surfaces corresponding to Na in the Na1 position show 
slightly larger $e_{g}$' hole pockets than when Na is in the Na2 position (shown in Fig. \ref{FS13}b).  
The position of Na varies the size of the hole pockets due to electrostatic interaction 
between Na$^+$ and the Co $a_{1g}$ orbitals. The $a_{1g}$ orbitals are shifted downward compared to the 
e$_g$' orbitals by positive Na$^+$ ions. This effect is stronger when Na is in the Na1 position versus the Na2 position because
the a$_{1g}$ orbital is elongated along the z-direction. The smaller shift of $a_{1g}$ downward when 
Na is in the Na2 position results in a smaller hole pocket \cite{PJMA}. 

\begin{figure}[tbp]
\includegraphics{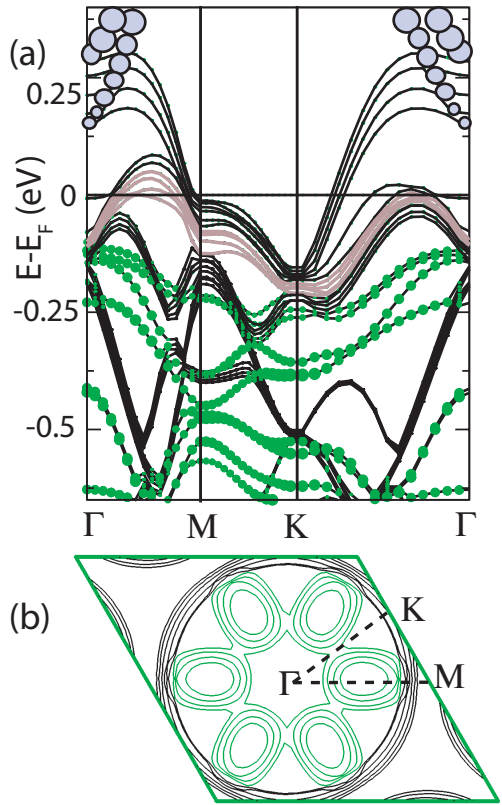}
\caption{Band structure for (a) H adsorbed on overstoichiometric surface
and Fermi surface of (b) H adsorbed on understoichiometric surface for Na$_{1/3}$CoO$_2$. 
Large and small circles indicate 
Na intercalant character and e$_g$' character, respectively.}
\label{Bands}
\end{figure}

The under- and over-stoichiometric compounds are, as noted previously, adjusted so that their electron count is equal to the 
stoichiometric one. Although the Fermi surfaces are 
mainly insensitive to this approximation, there is one notable difference that occurs due to the lack (under-stoichiometric) or 
surplus (over-stoichiometric) of Na on the surface itself.  Na, being a free-electron metal, features a parabolic, nearly 
free-electron band as a free atom or when it serves as an intercalant in a compound. It was recently pointed that the position of 
this band in intercalated graphites is crucial for superconductivity \cite{Csayni:2005}: in some compounds it is so low in energy 
that it crosses the Fermi level. It was observed that that the position of this intercalant band is quite sensitive to the 
interplanar distance \cite{Boeri:2007}. The interplanar distance determines how squeezed the Na electrons are between the host 
layers. Naturally, if the electrons are unbound on one side (surface), the intercalant band drops substantially (for CaC$_{6}$ 
this drop was found to be 0.5 eV \cite{calandra}). In the bulk, the Na band is high enough to be neglected 
when considering the Co-O bands, even at high Na concentrations (it sinks upon adding Na, for electrostatic reasons). On the 
surface, however, it is 4.21 eV lower than in the bulk for the overstoichiometric compound.  Fig. \ref{Bands}a shows the 
overstoichiometric bandstructure in which the Na band drops down below the unoccupied $e_g$' complex and overlaps with the 
unoccupied $a_{1g}$ bands near the $\Gamma$ point at approximately 250 meV above E$_F$.  For the stoichiometric compound (not 
shown), the bottom of the surface Na band is 670 meV higher than for the over-stoichiometric compound. Although the surface Na band 
does not cross E$_F$ for the low Na content ($x$=1/3) considered here, it does dip substantially beneath it for $x$=2/3, and 
therefore changes the fermiology significantly.  This may have relevance for the abrupt discontinuity in hole count compared to 
Luttinger's theorem noticed in systematic ARPES experiments.

The adsorption energy of a Na atom on the under-stoichiometric/stoichiometric surface is 3.92 eV/2.55 eV. These results indicate 
that it is highly favorable for Na to remain on the surface rather than desorb and that the cleaved surface of Na$_{1/3}$CoO$_2$ is 
highly susceptible to contamination by electron donor adsorbates. Although it seems unlikely that excess Na could be found to attach 
to the surface under UHV conditions, the presence of H$_2$O molecules or H atoms is much more conceivable. Even assuming that a 
perfectly clean oxide sample was placed inside a UHV chamber, a H$_2$O background and H contaminants are almost impossible to get 
rid of despite annealing at high temperatures and high vacuum pressures. In fact, it has been observed for some oxide surfaces that 
even after annealing to temps around 900K, H still appears to be stuck to the surface of the oxide \cite{Suzuki:2000}. We have 
calculated that a single H atom will adsorb to the surface of the over-stoichiometric compound with an energy of 2.84 eV. For the 
stoichiometric/under-stoichiometric compounds this increases to 3.50 eV /4.12 eV due to a reduction in Na-H repulsion. 
   
Surprisingly, the effect of adsorption of an H atom on the band structure and fermiology of Na$_{1/3}$CoO$_2$ is 
considerably more dramatic than simple electron donation would suggest.  As can be seen by comparing the highlighted 
$e_{g}$' surface bands of Fig. \ref{Bands} a to the bulk $e_{g}'$ bands of Fig. \ref{FS13}a, the presence of a H 
contaminant severely depresses the entire $e_{g}$' band complex.  The suppression is strongest along $\Gamma$-M where the 
bands formerly crossed E$_F$ without the contaminant, but the downshift in energy is obvious throughout the 
bandstructure.  The approximately -110 meV shift is congruent with what is seen in ARPES measurements 
\cite{yang05,qian06}, though 
no conclusive reports of surface contamination have yet been published.  Fig. \ref{Bands}a illustrates the effect of H on 
the over-stoichiometric surface while Fig. \ref{Bands}b shows the Fermi surface that results from H adsorption on the 
understoichiometric surface.  The suppression of the $e_g$' complex is weaker in the absence of Na on the surface, but 
still quite sufficient to push the $e_g$' pockets below E$_F$.

Although the Fermi surfaces in the stoichiometric unit cell are too complex to provide useful information, the position of the 
$e_g$' surface complex can nonetheless be gauged by examining the position of the $e_g'$ band center in comparison to the bulk. Note 
in Fig. \ref{Bands}a, that the part of the e$_g$' band complex that gives rise to Fermi surface hole pockets (along $\Gamma$-M) 
decreases in energy more than the band complex as a whole, so a strong shift in band center will clearly indicate suppression of 
these pockets. We find for the under-stoichiometric, stoichiometric and over-stoichiometric surface the band shifts are -42, -120 
and -137 meV versus the bulk (two CoO$_2$ layers beneath the surface). We have also checked that the bulk numbers are unaffected by 
H adsorption on the surface (or by the existence of the surface itself) by comparing third-layer band centers to full bulk 
calculation band centers.  Since the understoichiometric case exhibits the smallest band center shift and already has no surface 
$e_g$' hole pockets, these numbers demonstrate that the hole pockets are surely absent in the stoichiometric case.

The dramatic effects of H on the surface owe to two underlying physical effects.  The first is a simple filling of the Co-derived 
band complex due to electron donation from the adsorbed H.  This is, in reality, a very small effect, but since the e$_g$ pockets 
are already at the very top of the e$_g$' band complex and are the first part of the spectrum to fill upon electron addition, at 
higher Na content, even small amounts of H adsorbed on the surface could destroy the small Fermi surface pockets.  The second and 
dominant effect involves the positive Coulomb field created by ionized H just above the negative O ions which serves to exaggerate 
the main surface effect: O bands are strongly lowered, which decreases e$_g$' antibonding and destroys the pockets.  This effect is 
operative even if the adsorbate is H$^+$ rather than atomic H, as may be the case for dissociated water molecules in the vacuum.  
Because O anti-bonding with e$_g$' and a$_{1g}$ states is not uniform, there are also noticeable changes in the dispersion along 
$\Gamma$-M (See Figs. \ref{Bands}a and b).

\textit{Conclusions} Formation of a surface results in the $e_g$' hole pockets shrinking and possibly disappearing. This 
effect is highly dependent on Na ordering and content on the surface. Surface hydroxyl termination, which 
can occur through hydrogen contamination in UHV, results in complete suppression of 
the e$_g$' hole pockets. Our energetics indicate that once hydrogen is 
adsorbed it will be very difficult to remove. This is in good agreement with 
hydrogen adsorption studies on other oxide surfaces.  The Coulomb field of Na$^+$ or H$^+$ lowers the energy 
of the O electron bands which subsequently decreases Co-O bonding. Co-O hybridization increases 
along the $e_g$' symmetry parts of the band complex that result in hole pocket formation, relative to 
the $a_{1g}$ band complex. This results in $e_g$' hole pocket suppression. The decrease in size and absence  
of surface $e_g$' hole pockets is due to a combination of surface effects and surface contamination 
which agrees well with ARPES data.

We would like to thank C.S. Hellberg for helpful discussions. Research at NRL is 
funded by the Office of Naval Research. DP acknowledges support from the NRC Associateship Program.

\end{document}